\documentclass[prd,nofootinbib,superscriptaddress]{revtex4}

\usepackage{amsfonts,amssymb,amsthm,bbm,amsmath}

\usepackage[dvips]{graphicx}
\usepackage{psfrag}
\newcommand{\C}{\mathbb{C}}

\newcommand{\R}{\mathbb{R}}

\newcommand{\f}{\frac}

\newcommand{\SU}{\mathrm{SU}}
\newcommand{\SO}{\mathrm{SO}}

\newcommand{\Spin}{\mathrm{Spin}}

\def\la{\langle}
\def\ra{\rangle}

\newcommand{\cI}{{\mathcal I}}

\newcommand{\id}{\mathbb{I}}

\def\q{(d_{j_0}, \Theta)}
\newcommand{\be}{\begin{equation}}
\newcommand{\ee}{\end{equation}}
\newcommand{\bes}{\begin{eqnarray}}
\newcommand{\ees}{\end{eqnarray}}

\newcommand{\Ref}[1]{(\ref{#1})}

\def\tj{{\{10j\}}}

\def\pp{\partial}

\newtheorem{theorem}{Theorem}[section]

\newtheorem{proposition}[theorem]{Proposition}

\newcommand{\tabl} [2] {\begin{array} {#1} #2 \end{array}}

\begin{document}

\title{\large\bf Boundary State Stability under Spinfoam Evolution for the Quantum 4-Simplex}

\author{{\bf Ma\"\i t\'e Dupuis}}\email{maite.dupuis@ens-lyon.fr}
\affiliation{Laboratoire de Physique, ENS Lyon, CNRS-UMR 5672, 46 All\'ee d'Italie, Lyon 69007, France}

\author{{\bf Etera R. Livine}}\email{etera.livine@ens-lyon.fr}
\affiliation{Laboratoire de Physique, ENS Lyon, CNRS-UMR 5672, 46 All\'ee d'Italie, Lyon 69007, France}

\date{\small \today}

\begin{abstract}

In the spinfoam framework for quantum gravity, we investigate the conditions to have a physical quantum state for the Barrett-Crane model for the 4d quantum gravity path integral.
More precisely, we look at boundary states stable under evolution with respect to the spinfoam kernel.
Hee, we look at the simplest case of a single 4-simplex boundary and show that the requirement of working with a physical boundary state fixes the width of the semi-classical Gaussian wave-packet for the boundary 3d geometry. This is directly relevant to the graviton propagator calculations done in this framework, since the Gaussian width enters the numerical factors in front of the graviton correlations in the large scale asymptotical limit. Finally, we discuss the application of our computations to the Barrett-Crane model beyond the first order (of a single 4-simplex in the bulk) and to the more recent EPRL-FK spinfoam model.

\end{abstract}

\maketitle

\section*{Introduction}

Spinfoam models provide a background independent framework for a regularized path integral for (loop) quantum gravity. They define transition amplitudes between quantum states of geometry through state-sum models which can be understood as discrete space-time geometries. In this context, a recent development has been a proposal for reconstructing the graviton propagator in this discrete setting from correlations between geometrical observables such as the areas of elementary surfaces \cite{graviton}. This allows to extract some semi-classical correlations at large scales, which we hope to compare with the perturbative calculations performed in quantum general relativity treated as a quantum field theory.

Since the original proposal, they has been a lot of work developing this line of research, mainly focusing on creating the new mathematical tools needed for these semi-classical calculations and on using these computations as a criteria to select spinfoam models with a correct semi-classical behavior and discriminate between them (see e.g \cite{alesci}). All these developments hint towards the fact that the graviton propagator in spinfoam models lead back to Newton's law for gravity at large scales while being regularized at the Planck scale. This behavior has been confirmed by numerical simulations in the simplest cases \cite{numeric1,numeric2}.

The main ingredients of these ``spinfoam graviton propagator" calculations are: a suitable boundary state peaked on a classical 3-geometry, the spinfoam amplitudes for the bulk geometry and relevant observables probing the space-time geometry.
The main spinfoam models used to define the bulk amplitudes are the Barrett-Crane model, which exists in both its Euclidean version \cite{bc1,bubblediv} and its Lorentzian counterpart \cite{bc2,bc2bis}, and the more recent EPRL-FK models \cite{epr,eprl,fk,ls1,ls2}
and their generalizations \cite{eprl,polish}.

Considering boundary states, while most of the recent work has focused on building quantum coherent states with good semi-classical properties, one important issue is the requirement that the states are physical i.e solve the Hamiltonian constraint on the 3d boundary.
As we will review and discuss later, this ``physical state" criteria can be entirely formulated in term of a compatibility equation between the boundary state and the spinfoam bulk amplitude, which translates the requirement of stability of the boundary state under the evolution defined by the spinfoam model.
This question was investigated in the framework of the Ponzano-Regge spinfoam model for 3d quantum gravity, more particularly in a toy model where the three-dimensional space-time is triangulated by a single tetrahedron \cite{toymodel3d}. In this case, the correlations between length observables can be thoroughly studied both analytically and numerically \cite{graviton3d}. In this context, the requirement of working with a physical boundary state was shown to determine explicitly and uniquely the width of the gaussian wave-packet defining the state \cite{physical3d}. This width is relevant in the context of the geometrical correlations because it enters the exact numerical factor in front of these correlations. Therefore, if we would like to have the exact correlations and not only their scaling properties, we need a definite prediction of that width.

\smallskip

In the present paper, we investigate the consequences of the physical state requirement for the Euclidean Barrett-Crane model for the simplest case of a space-time triangulation constructed from a single 4-simplex. In this context, we show that this requirement fixes uniquely the width of the quantum boundary state (in term of the classical data) similarly to what happens in the 3d toy model. In the next section, we will introduce and quickly review the framework of the spinfoam graviton propagator. Then in section II, we will review the definition of the (Euclidean) Barrett-Crane spinfoam model. Section III presents our results about identifying suitable physical boundary states for the Barrett-Crane model.
We consider both the simplest cases of a decoupled Gaussian ansatz and the more interesting case of a coupled Gaussian ansatz involving the Hessian of the Regge action (for a single 4-simplex).
Finally, we will conclude by a short discussion on the relevance of our physical states under renormalization of the Barrett-Crane model and on the possible extension of our results to other spinfoam models.

\smallskip

Before proceeding to the analysis of the Barrett-Crane spinfoam amplitude for a 4-simplex, we would like to discuss to which extent our notion of boundary states stable under spinfoam evolution can be considered as an actual physical state. From the strict perspective of the canonical framework, a physical quantum state should solve the Hamiltonian constraint (and actually also solve the vector constraints generating space diffeomorphisms). Thinking of this requirement in classical terms, this means that the 3-metric on our boundary should be physical, i.e satisfy the constraints, and that we should furthermore consider equivalence classes of boundary 3-metrics under the gauge transformation(s) generated by the constraint(s). In our spinfoam framework, considering the 3d case with a single tetrahedron in the Ponzano-Regge model, there is a single truly physical state from the canonical point of view, which solves all the constraints. This is the completely flat boundary state, which is given by the 6j-symbol. The procedure followed in \cite{physical3d} is to look for semi-classical states on the same tetrahedron, which are peaked around a certain classical 3-metric satisfying the constraints (here defined as the lengths of the tetrahedron's edges) and which are stable under evolution defined by the Ponzano-Regge spinfoam amplitude. The requirement of being fixing the tetrahedron's edge lengths amounts to a gauge-fixing of the  Hamiltonian (and diffeomorphism) constraint(s). This allows us to identify a single quantum boundary state for each classically admissible edge-length configurations of the tetrahedron \cite{physical3d}. Each of these quantum boundary states describe a flat physical 3d-metric within the tetrahedron. Thus we refer to these boundary states as physical in the spinfoam context, although they are not strictly physical states from the canonical viewpoint.

We follow the same logic in the present study of the quantum 4-simplex in the Barrett-Crane model. The basic spinfoam amplitude for the 4-simplex implies that the bulk 4-metric is flat, although we will discuss in the conclusion the possibility of effective spinfoam amplitudes for the 4-simplex inducing curvature. Thus we expect once again a unique physical boundary state for 4-simplex from the strict canonical theory definition. Nevertheless, we will look for semi-classical states which are peaked on fixed values of the area of the boundary triangles and which are stable with respect to the Barrett-Crane spinfoam amplitude. Fixing the triangle areas amounts again to a partial gauge-fixing of the Hamiltonian and diffeomorphism constraints. And we then expect a unique such stable semi-classical boundary state for each sets of classically admissible areas. These can be considered as particular gauge-fixings of the unique canonical physical flat state. By misuse of language, we shall nevertheless refer, in the  specific spinfoam context, to our quantum boundary states stable under spinfoam evolution as physical states.

\section{Spinfoams and the ``Graviton Propagator" Framework}

The spin foam approach to quantum gravity provides a path integral formulation for the canonical framework of loop quantum gravity (LQG).
It is based on a reformulation of general relativity of an ``almost-topological" gauge field theory and the spinfoam path integral is constructed as a discretization of the continuum path integral. Without focusing especially on the specific case of quantum gravity, we describe below the generic features of the spinfoam framework for an arbitrary gauge group $G$.
As it is well-known, the Hilbert space of quantum geometry states for loop quantum gravity  is spanned by spin network states $|\psi \ra$ where  in $\psi=(\gamma, j_l, i_n)$: $\gamma$ is a graph, $j_l$ is a ``spin" labeling an irreducible representation of the gauge group $G$ associated to the link $l$ of the graph, and $i_n$ is associated to the node $n$ of $\gamma$ and labels  intertwiners. We consider a 4d space-time region $\mathcal{M}$  with a 3d boundary $\Sigma$. The spin network state $|\psi\ra$ defines the quantum state of geometry of the boundary $\Sigma$, then the spin foam amplitudes define the dynamical probability amplitude of that state and are supposed to contain the whole dynamical content of quantum gravity. More precisely, the standard ansatz for local spinfoam amplitudes can be presented in the following general form:
\be \label{propaSF}
K[\psi]=\sum_{C|\partial C= \gamma_\psi} w(C) \sum_{j_f, i_e} \prod_f A_f(j_f) \prod_e A_e(j_f, i_e) \prod_vA_v(j_f, i_e)
\ee
where the sum is taken over two-complexes $C$ and their ``coloring"  $c\,\equiv\,\{j_f, i_e\}$. The 2-complexes are constrained to fit the graph $\gamma_\psi$ of the spin network $\psi$ at the boundary.
The representations $j_f$ are associated to the faces $f$ of $C$ and the intertwiners $i_e$ to its edges $e$. They are also constrained to fit the representations $j_l$ and intertwiners $i_n$ of the state $\psi$ living on the boundary.
Then a spinfoam is defined as a colored two-complex, namely a couple $(C, c)$.
Finally, the spinfoam amplitude is made of four types of factors. First, $w(C)$ is a statistical weight that depends only on the two-complex $C$ (similar to the symmetry factor for Feynman diagrams in standard quantum field theory). $A_f$ are weight factors  associated to the faces and $A_e$ are amplitudes associated to the edges of the 2-complex. These three types of factors can be interpreted as measure factors of the discrete path integral defined by $K[\psi]$. All the dynamical information is encoded in the vertex amplitude, $A_v(j_f, i_e)$, which is an amplitude associated with each vertex $v$ of the two-complex $C$ and depends on the spins $j_f$ and intertwiners $i_e$ living on the faces and edges around that vertex.

The amplitudes are usually assumed to be local, that is they depend only on the coloring of adjacent simplicial elements. Thus, $A_f$ is a function of the representations located on the face $f$, $A_e$ is a function of the intertwiner assigned to $e$ and the representations on the faces containing $e$, whereas $A_v$  depends on the representations on the faces and on the intertwiners on the edges containing the vertex $v$.
Generally speaking, the choice of the vertex amplitude $A_v$ corresponds to the choice of a specific form of the hamiltonian operator in the canonical theory.

Since a spin network state defines a quantum state for the boundary geometry, spin foams are thus interpreted as representing the states of the bulk defining the quantum space-time interpolating between given boundary data. This  can be clearer when considering an initial spin network $\psi$ and a final spin network $\psi^\prime$. Then a spin foam $\mathcal{F}: \psi \rightarrow \psi^\prime$ defined by the 2-complex $C$ bordered by the supporting graphs of $\psi$ and  $\psi^\prime$, respectively $\gamma$ and $\gamma^\prime$, represents a transition from the spin network state $\psi=(\gamma, j_l, i_n)$ into $\psi^\prime=(\gamma^\prime, j_l^\prime, i_n^\prime)$. Nodes and links in the initial spin network $|\psi\ra$ evolve into 1-dimensional edges and faces. New links are created and spins are reassigned at vertexes. This defines a foam-like structure whose components inherit the spin representations from the initial spin network $\psi$ and are at the end compatible with the spin representations of the final spin network $\psi^\prime$. The propagator kernel $K[\psi,\psi^\prime]$ is obtained summing over all the spinfoams compatible with the boundary data:
\be \label{propaSFcompact}
K[\psi,\psi^\prime]=\sum_{\mathcal{F}|_{\partial \mathcal{F}=\psi \cup \psi^\prime }} \mathcal{A}_\mathcal{F}[\psi,\psi^\prime]
\ee
where we have introduced here a more compact notation than in the previous formula (\ref{propaSF}): the sum over the compatible spin foam $\mathcal{F}$ gathers the sum over the two-complexes $C$ compatible with the graphs $\gamma$ and $\gamma^\prime$ and the sums over the G-representations compatible with the representations $j_l$ and $j_l^\prime$ and over the intertwiners compatible with $i_n$ and $i_n^\prime$.  $\mathcal{A}_{\mathcal{F}}[\psi,\psi^\prime]$ is the spin foam amplitude associated to the spin foam $\mathcal{F}: \psi \rightarrow \psi^\prime$ interpolating between the initial and final states. It is given as above as a product of vertex, edge and face amplitudes: $A_v, \, A_e, \, A_f$. This special case where the boundary is made of two disconnected pieces, which we can interpret as the initial boundary and the final boundary, allows a clearer connection with the canonical framework, whose goal is to define transition amplitudes between initial and final states.

In the present work, we will consider the case where the boundary is connected and the kernel $K$ is then defined as a function of only one boundary spin network $\psi$:
\be
K[\psi]=\sum_{\mathcal{F}|_{\partial\mathcal{F}=\psi }} \mathcal{A}_\mathcal{F}[\psi]
\ee
where we used the same compact notation as above.

Before going further, we have to emphasize that the sum over spinfoam $\mathcal{F}=(C,c)$ is not exactly well-defined. Usually, for fixed $C$, the sum over colorings $c$ is well-controlled. On the other hand, controlling the full sum over two-complexes is a much more subtle issue. It is nevertheless non-perturbatively defined as quantum field correlations  in the context of group field theory (see e.g. \cite{GFT}). In this context, $\mathcal{A}_\mathcal{F}$ actually  depends on the GFT coupling constant $\lambda$. Indeed the statistical weight $w(C)$ for a two-complex $C$ is given by the symmetry factor of the two-complex (considered as a Feynman diagram for the GFT) times a factor $\lambda^V$ where $V$ is the number of vertices of the two-complex. Then the sum over 4-geometries in $K[\psi]$ can be written as a power series in $\lambda$; that is:
$$
K[\psi]=\sum_{V=0}^\infty \lambda^{V} K_V[\psi]
$$
where $K_V[\psi]=\sum_{\mathcal{F}^V|_{\partial \mathcal{F}^V=\psi }} \mathcal{A}^V_\mathcal{F}[\psi]$ is the sum over spinfoams $\mathcal{F}^V$ with $V$ vertices. Assuming that the perturbative expansion in $\lambda$ is well-defined and that we are working with a suitable semi-classical boundary state $\psi$, we expect that the dominant contributions come from the simplest spinfoams and we consider the lowest order terms in the expansion of $K[\psi]$ in $\lambda$ as giving the leading order for the path integral and correlations.


\medskip

Let us now describe the spinfoam framework for deriving the graviton propagator from correlations between area observables.
We start by considering a semi-classical spin network functional $\Psi_{q}[\psi]$ peaked on a classical 3d metric $q$ for the boundary $\Sigma$. Let us remind the reader that $\psi=(\gamma,j_l,i_n)$ labels basis states of the Hilbert space of spin networks. We further require that this boundary state $\Psi_{q}[\psi]$ induces a space-time structure in the bulk peaked around the flat Minkowski metric. In particular, this normally fixes the classical boundary $q$ to be the 3-metric induced on $\Sigma$ by the Minkowski metric on ${\cal M}$. Then we construct correlations between the metric fluctuations for the chosen boundary state $\Psi_{q}$:
\be \label{twopointSF}
W^{abcd}(x,y;q)=\sum_{\psi}K[\psi] \la \psi | h^{ab}(x) h^{cd}(y) |\psi \ra \Psi_{q}[\psi]
\,=\,{\textrm Tr}\, \left[K \,h^{ab}(x) h^{cd}(y)\,\Psi_q\right]\,,
\ee
where the trace is taken over the Hilbert of spin networks. Here $x$ and $y$ are two points localized on the boundary $\Sigma$. In our discrete spin network setting, they are usually determined as nodes of the graph $\gamma$ underlying the spin network state $\Psi_q$. The metric fluctuations $h^{ab}(x) h^{cd}(y)$ are usually constructed as geometrical quantities. They are two basic types of such geometric observables in the discretized geometry setting of spinfoams: the diagonal components of the metric tensor can be interpreted as areas and the off-diagonal components as dihedral angles between simplices.

Finally, this formula defines the 2-point function for the gravitational field in the spinfoam framework. It can be considered as the equivalent of the standard 2-point function of the conventional quantum field theory  framework, which defines the graviton propagator \cite{graviton}:
\be \label{2point}
W_{\mu \nu \rho \sigma} (x, y)= \la 0 | T\{ h_{\mu \nu} (x) h_{\rho \sigma} (y) \}| 0\ra.
\ee

In the spinfoam setting,  $\Psi_{q}[\psi]$ is a semi-classical state peaked on both intrinsic and extrinsic geometry \cite{graviton}.
This boundary data will determine the 4-metric induced in the bulk.
%
Moreover, it should be a physical state, that is a solution a the Hamiltonian constraint. This ``physical state criteria" can be formulated by the two following conditions: 
\be \label{conditionsSF} \left\{ \tabl{l}{
\sum_\psi |\Psi_q[\psi]|^2=1 \\
\sum_\psi K[\psi] \Psi_q[\psi]=1
} \right.
\ee
The first condition is the normalization of the boundary state, while the second condition translates truthfully the requirement to work with a physical state and can be considered as the ``Wheeler-deWitt" condition. It is the consequences of these conditions on $\Psi_q$ that we are going to study.

\section{The Barrett-Crane Spinfoam Model}
We now focus  on a specific spinfoam model: the Barrett-Crane model of 4d Riemannian Quantum Gravity \cite{bc1}.

\subsection{Simple Spin Networks and the Barrett-Crane Amplitudes}
The spin foam amplitude for the Euclidean Barrett-Crane model  is given by the evaluation of a spin network $\psi$ with group $\Spin(4)$. $\Spin(4)$ is the double cover of $\SO(4)$. Using the homomorphism $\Spin(4)=\SU(2)_L\times\SU(2)_R$ the irreducible representations (irreps) of $\Spin(4)$ are labelled by two halh-integers: $\cI=(j_L,j_R)$ corresponding to the irreps of the two $\SU(2)$ sectors and group elements decompose as the product of two left and right rotations $G=g_Lg_R$. In the Barrett-Crane model, each node $n$ of the graph $\gamma$ of $\psi$ is associated to a $\Spin(4)$ group element $G_n$ and each link $l$ is labelled by a simple irreps of $\Spin(4)$ which corresponds to a couple of half-integers $\cI_l=(j_l,j_l)$: the SU(2) irreps are the same in the left and right sectors. The evaluation then reads \cite{barrett}
\be
\mathcal{A}[\psi] \equiv \int_{\Spin(4)} \prod_n dG_n \prod_i
\mathcal{K}_{\mathcal{I}_i} (G^{-1}_{s(i)}G_{t(i)}),
\ee
where we use the Haar measure $dG=dg_L dg_R$.
%
%
The kernel $\mathcal{K}_{\mathcal{I}}(G)$ is the matrix element of $G$ on the $\SU(2)$-invariant vector $|\cI, 0\ra$ in the $\cI$ representation with $\cI=(j, j)$ a simple representation of $\Spin(4)$. Here $\SU(2)$ is the diagonal rotation group, corresponding to the subgroup of 3d rotations.
Expressing the invariant vector in term of left/right components
\be
|\cI, 0\ra = \f{1}{\sqrt{d_j}} \sum_m (-1)^{j-m} |j, m\ra_L |j, -m\ra_R
\ee
where $d_j=2j+1$ is the dimension of the $\SU(2)$ representation of spin $j$. It is straightforward to show that the kernel $\mathcal{K}_{\mathcal{I}}$ is simply given by  the $\SU(2)$ character $\chi_j(g)$, defined as the trace of the group element in the $j$-representation of $\SU(2)$.
Parameterizing the $\SU(2)$ group elements as
\be
g(\theta, \hat{n})= \cos(\theta) \id + i \sin(\theta) \hat{n} \cdot \vec{\sigma}, \quad \theta \in [0, \pi],
\ee
the characters depend entirely on the class angle $\theta$ (half the rotation angle) and are expressed as:
\be
\chi_j(g)=\f{\sin d_j \theta}{\sin \theta}.
\ee
Then using the properties of invariance of the Haar measure, it can be proved that the relativistic spin network evaluation is actually a 3d object involving only integrals over $\SU(2)$ \cite{simone1}:
\be
\mathcal{A}= \int_{\SU(2)} \prod_n dg_n \prod_i \f{1}{d_{j_i}} \chi_{j_i}(g^{-1}_{s(i)} g_{t(i)}).
\label{evaluation}
\ee
This spin network evaluation then defines the vertex amplitude for the Barrett-Crane model.

The Barrett-Crane model was originally derived for  specific two-complexes $C$, which are the dual 2-skeleton of 4-dimensional triangulations $\Delta_4$.
%
%
Starting with the triangulation  $\Delta_4$, the dual two-skeleton $C$ is constructed  by associating to each simplex a point in its interior. An edge of the skeleton connects two points, which corresponds to two 4-simplices having a common tetrahedron. Then, a triangle of the triangulation $\Delta_4$ correspond to a face of the two-complex $C$ formed by the edges of the dual tetrahedra having this triangle in common. This duality allows to derive the spin foam amplitudes from a quantization of the geometry of a 4-simplex.
%
%
Indeed, the amplitude for a single 4-simplex defines the vertex amplitude of the spinfoam model. Arbitrary 2-complexes are constructed by gluing 4-simplices together and the corresponding spinfoam amplitudes are given by the product of these vertex amplitudes corresponding to each 4-simplex.

\begin{figure}[ht]
\begin{center}
\includegraphics[width=5cm]{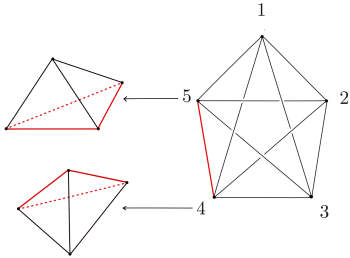}
\end{center}
\caption{The 4-simplex (or pentahedral) boundary spin network. We label the nodes $n= 1, . . . 5$. In the dual picture, they are in correspondence with tetrahedra of the boundary triangulation. Two of them are represented. The ten links $nm$, on the other hand, are dual to triangles. Consider for instance the link 45: this is dual to the triangle shared by the tetrahedra 4 and 5. Associated with the link 45 is the dihedral angle $\theta_{45}$ between the tetrahedra 4 and 5.} \label{4simplexPlusTetra}
\end{figure}

Considering a single 4-simplex, the vertex amplitude is constructed as the evaluation of its boundary spin network. The boundary graph is constructed as the dual of the boundary of the 4-simplex: each tetrahedron is mapped on a node of the graph and each triangle shared by two tetrahedra is mapped to a link between these nodes. This gives a graph $s$ with 5 nodes $n \in \{1, \cdots, 5\}$ and 10 links between them, as represented in fig. \ref{4simplexPlusTetra}.
%
%
Then the vertex amplitude $\mathcal{A}[s]$ is given up to normalization factors by the evaluation of the boundary spin networks and defines the Barrett-Crane $\{10j\}$-symbol:
\be \label{10j}
\{10j\}\equiv  \int_{\SU(2)} \prod_{n=1}^{5} dg_n \prod_{i=1}^{10} \chi_{j_i}(g^{-1}_{s(i)} g_{t(i)}),
\ee
following equation (\ref{evaluation}).
The normalization ambiguity leads to an ambiguity in the definition of the Barrett-Crane model. It corresponds to arbitrary gluing factors between 4-simplices, or equivalently to ambiguity in defining the spinfoam edge amplitude ${\cal A}_e$. We will come back to this issue in the next section.

\subsection{The 4-Simplex and the Boundary States}
We now focus on the simplest 4d boundary spin network: a pentahedral that we denote $s=(\Gamma, j_{ab}, i_a)$ which is the dual of a single 4-simplex. We have  introduced  the notation $(a\, b)$ for $i$ linking the nodes $a$ and $b$.  $a, b=1\cdots 5$ label the five nodes of the pentahedral; in the triangulation picture, there are in correspondence with tetrahedra of the boundary  4-simplex and links $i\equiv (a\, b)$ are dual to triangles. Thus, in the triangulation picture,  while the five group elements $g_a \, \in$ $\SU(2)$ are associated to the 5 tetrahedra of the 4-simplex, the ten representations $j_i \equiv j_{ab}$ for $a\ne b$ can be seen as attached to the triangles of the 4-simplex.
The amplitude $\mathcal{A}_\mathcal{F}$ of a spin foam $\mathcal{F}\equiv(C,c)$ where the two-complex $C$ is bordered by the boundary 4-simplex graph $\Gamma$ and its color $c$ is compatible with $\{j_{ab}, i_{a}\}$ is then given (up to normalization factors) by:
\be \label{SFamplitude}
\mathcal{A}[s]= \prod_{a<b} (d_{j_{ab}})^2 \prod_{a<b}\left(A_e(d_{j_{ab}})\right)^{1/2}\tj
\ee
where  $A_e(j_f)$ is the edge amplitude left unspecified here and the $\tj$-symbol, the vertex amplitude of the spin foam, defined by (\ref{10j}). Let us point out that $\mathcal{A}[s]=K[s]$ in the first order in the group field theory coupling constant $\lambda$ in the specific case of the pentahedral boundary spin network $s$. We now focus on the $\tj$-symbol which is defined by the following group integral: 
\be \label{10jsymbol}
\tj\,=\,
\int_{\SU(2)} [dg_a]^{\otimes 5}\,
\prod_{a<b}\chi_{j_{ab}}(g_ag_b^{-1})
\ee
The $\{10j\}$ symbol admits a geometrical interpretation associated to the 4-simplex. The key fact is that it can be written as an integral over ten class angles $\theta_{ab} \in [0, \pi]$ of the group elements $g^{-1}_{a}g_{b}$:  $\theta_{ab}$ is the dihedral angle and we use the convention $\theta_{aa}=0$. In the triangulation picture, it is the angle between the outward normals to the tetrahedron $a$ and the tetrahedron $b$ of the 4-simplex. The $\{10j\}$ symbol then reads,
\be
\{10j\}=\int d\mu[\theta_{ab}] \prod_{a<b} \chi_{j_{ab}}(\theta_{ab}),
\ee
where the measure on the ten class angles $d\mu[\theta_{ab}]$ takes into account that the group elements $g^{-1}_{a}g_b$ are obviously not independent. The measure is in fact simply given by a constraint \cite{6jsaddle2}
\be
d\mu[\theta_{ab}] = \prod_{a<b} d\theta_{ab} \sin \theta_{ab} \delta \left(\det G_{ab} \right)
\ee
$G$ is the Gram matrix, a symmetric $5\times 5$ matrix defined as $G_{ab} = \cos \theta_{ab}$. This constraint $\delta(G)$ contains all the geometric information and allows to relate the $\{10j\}$-symbol to the Regge amplitude (which describes discretized gravity) for a Euclidean 4-simplex in the large $j$ asymptotics \cite{6jsaddle2, simone1}. 
The independent variables are the areas of the triangles of the 4-simplex representing the triangulation or equivalently the dimension of the representation $j_{ab}$, $d_{j_{ab}}=2j_{ab}+1$, since $A_{ab}=l_P^2d_{j_{ab}}$. The geometry of the 4-simplex is fixed by given the ten areas $A_{ab}=d_{j_{ab}}$ and the dihedral angles can be considered as functions $\theta_{ab}(d_{j_{cd}})$ of the areas. Actually, we scale the area $d_{j_{ab}}$ instead of the spins $j_{ab}$, then the asymptotic behavior on a single 4-simplex at the leading order\footnote{The fact to scale the amplitude with the irrep dimension $d_{j_{ab}}$ instead of the spin $j_{ab}$ has no effect on the leading order.} is given by:
\be \label{asympt}
\tj \, \sim \, P(d_{j_{ab}}) \cos \left( \f{1}{l_P^2}S_R[d_{j_{ab}}]\right) + D(d_{j_{ab}})
\ee
where $S_R[d_{j_{ab}}]= \sum_{a<b}d_{j_{ab}}\theta_{ab}$. The function $P(d_{j_{ab}})$ is a slowly varying factor, that grows as $\zeta^{-9/2}$ when scaling all triangle areas $d_{j_{ab}}$ by $\zeta$.
$D(d_{j_{ab}})$ is a contribution coming from degenerate configurations of the 4-simplex. This is a non-oscillating term which has no geometrical meaning. It was found to scale like $1/\zeta^2$ and thus dominate the large spin limit of the $\tj$-symbol. Moreover, it was shown in a recent paper \cite{numeric2} that even after having removing the dominating contribution of order $1/\zeta^2$, there are additional non-oscillating contributions still hiding the Regge term. However, this sick term is negligible in most computations we are interested in, such as in the computation of  (\ref{twopointSF}) (see \cite{graviton}) or as we will see in the computation of (\ref{conditionsSF}), because it will not match the boundary data induced by $\Psi_q$ which peaks the asymptotic around the non-degenerate semi-classical configuration.

On the other hand, the unspecified edge amplitude $A_e$ in (\ref{SFamplitude})  reflects the normalization ambiguity of the $\tj$-symbol and different choices leads to different versions of the Barrett-Crane model; we will specify this normalization factor latter.
%

Since in the Barrett-Crane model, the vertex amplitude $A_v$, defined by (\ref{10jsymbol}),
and therefore the propagator kernel $K[s]=\mathcal{A}[d_{j_{ab}}]$, given by (\ref{SFamplitude}), are independent of the intertwiners, we  can  rewrite the two conditions (\ref{conditionsSF}), restricting $\psi_q$ to be a physical state,  as\footnote{It is often useful for spinfoam asymptotic calculations to do a Fourier transform and go back to group variables (see e.g. \cite{simone1} for the application to the spinfoam graviton propagator). In the present case, this is straightforward to do if we choose a factorizable ansatz, i.e. a product state of the form $\psi(j_1,..,j_{10})=\prod_{i=1}^{10} \phi_i(j_i)$. We further simplify it by taking an equilateral ansatz, i.e $\phi_i(j)=\phi(j)$ for all triangles $i=1..10$. Then following \cite{simone1}, we introduce the Fourier transform of the state:
$$
f(g)=\sum_j \phi(j) \chi_j(g).
$$
Then the two conditions for a physical state simply translates to:
$$
\int dg\, |f(g)|^2=1,
$$
$$
\int [dg_m]^5\,\prod_{a<b} f(g_mg_n^{-1})=
\int [d\theta_{mn}]^{10}\delta(\det [\cos\theta_{mn}])
\prod_{m<n}f(\theta_{mn})\,=\,1.
$$
}:
\be \label{norm}
\sum_{j_{ab}} |\psi_q(d_{j_{ab}})|^2=1 \quad \textrm{ for the normalized condition,}
\ee
\be \label{dyna}
\sum_{j_{ab}} K(d_{j_{ab}})\psi(d_{j_{ab}})=1 \quad \textrm{ for the "Wheeler-deWitt" condition,}
\ee
with the propagator kernel a function of the dimensions $d_{j_{ab}}$ only
\be \label{kernel}
K[j_{ab}]= \mu \left(\prod_{a<b}d_{j_{ab}}\right)^{\sigma}\tj
\ee
with $\mu$ and $\sigma$ undetermined coefficients. The edge amplitude $A_e$ is taken into account in this factor $\mu \left(\prod_{a<b}d_{j_{ab}}\right)^{\sigma}$.

For more generic spinfoam models such as the EPRL-FK models, we need to take into account the intertwiners in the definition of the spinfoam vertex and of the physical states. This is postponed to a further work.

\section{Looking for Physical States}
In the following, we will  work on the Barrett-Crane model in the large spin limit; that is we will consider the asymptotic formula of the $\tj$-symbol (\ref{asympt}) in the propagator kernel formula (\ref{kernel}). 
We now focus on the issue of the semi-classical boundary state. The function $\Psi_q$ should describe the boundary value of the gravitational field on the boundary 4-simplex. We thus consider a state peaked on the geometry of a regular 4-simplex: $q=(d_{j_0}, \Theta)$. The simplest possibility is to choose a Gaussian peaked on theses values:
\be \label{gaussian}
\Psi_{\q}(d_{j_{ab}})= e^{-\sum_{a<b,c<d} \alpha_{cd}^{ab}(d_{j_{ab}}-d_{j_0})(d_{j_{cd}}-d_{j_0})+i\sum_{a<b} \Theta \, d_{j_{ab}}}
\ee
The phase of this semi-classical state determines where the state is peaked in the conjugate variables: $\Theta$ is the variable conjugate to the spin $j_0$ and it codes the extrinsic geometry of the boundary. $\alpha_{cd}^{ab}$ is a given ten by ten matrix. It depends on $d_{j_0}$ in such a way that the relative uncertainties of area and angle on this state become small in the large $d_{j_0}$ limit, namely:
\be
\f{\la \Psi_q|\Delta d_{j_{ab}}| \Psi_q\ra}{\la\Psi_q| d_{j_{ab}}| \Psi_q\ra} \rightarrow 0, \qquad \qquad \f{\la\Psi_q|\Delta \theta_{ab}| \Psi_q\ra}{\la\Psi_q| \theta_{ab}| \Psi_q\ra} \rightarrow 0 \qquad \qquad \forall \,a<b
\ee
Assuming  that the matrix elements $\alpha_{cd}^{ab} \sim \alpha d_{j_0}^{-n}$ in the large spin limit with $\alpha$ which does not scale with $d_{j_0}$, the fluctuation determined by the gaussian state (\ref{gaussian}) are of the order:
\be \label{uncertainties}
\f{\la\Psi_q|\Delta d_{j_{ab}}| \Psi_q \ra}{\la\Psi_q| d_{j_{ab}}| \Psi_q \ra} \sim \f{d_{j_0}^{n/2-1}}{\sqrt{\alpha}}, \qquad \qquad \f{\la\Psi_q|\Delta \theta_{ab}| \Psi_q \ra}{\la \Psi_q| \theta_{ab}| \psi_q \ra} \sim d_{j_0}^{-n/2}\sqrt{\alpha}
\ee
which restricts $n \in  ] 0, 2[$. \\
In the following, we thus focus on a Gaussian state as boundary state such as the matrix elements of the ten by ten matrix $\alpha^{ab}_{cd}$ are given by
\be
\alpha^{ab}_{cd} \sim \f{\alpha}{d_{j_0}^n}= \f{a+ib}{d_{j_0}^n} \quad \textrm{ with } n \in ]0,2[ \textrm{ and } a, b \in \R
\ee
in the large spin limit. We now study the consequences of the two conditions (\ref{norm}) and (\ref{dyna}) on this boundary state. That is the aim is now to determine the consequences on $\alpha_{cd}^{ab}$ of the requirement that our boundary state is a physical state.

\subsection{Semi-Classical States: the Decoupled Gaussian Ansatz}
We first start by considering an additional ansatz for the boundary state. We take a factorized boundary state:
\be
\Psi_q[j_{ab}]= \prod_{a<b}\phi(j_{ab})
\ee
where each $\phi(j_{ab})$ is peaked around the background value $q=(d_{j_0}, \Theta)$ which corresponds to an equilateral 4-simplex.
%
In this simplest case, the ten by ten matrix $\alpha_{cd}^{ab}$ reduces to a diagonal matrix $\alpha \id_{10\times 10}$.
Such a factorized boundary state has been used in  \cite{numeric2} since it is up to now the only setting in which numerical simulations can be performed. However, this assumption has not been tested yet. Could such a decoupled gaussian state capture a true physical state? Could it satisfy conditions (\ref{norm}) and (\ref{dyna})?

Two choices for a factorized boundary state have so far appeared in the literature: \begin{itemize}
\item  the Gaussian state \cite{graviton, numeric1}, where each factor is given by,
\be \label{gaussianf}
\phi(j)= e^{-\alpha (d_{j}-d_{j_0})^2}e^{i\Theta d_{j}}
\ee
where $\alpha$ is a complex number.
\item The Bessel-based state \cite{simone1, graviton3d}, where each factor is given by:
\be \label{bessel}
\phi_B(j)= \f{I_{|d_j-d_{j_0}|}(\f{d_{j_0}}{\alpha})-I_{d_j+d_{j_0}}(\f{d_{j_0}}{\alpha})}{\sqrt{I_0(\f{2d_{j_0}}{\alpha})-I_{2d_{j_0}+1}(\f{2d_{j_0}}{\alpha})}} \cos\left(d_j\Theta\right)
\ee
\end{itemize}
In the large spin limit regime, we focus only on the Gaussian ansatz. Indeed, in the large spin limit the Bessel part of (\ref{bessel}) reduces to the Gaussian in (\ref{gaussianf}). Therefore, at the leading order, the only difference between (\ref{gaussianf}) and (\ref{bessel}) is in the phase: the phase in (\ref{gaussianf}) is complex whereas the phase in (\ref{bessel}) is real. A gaussian state with a real phase would lead to the same results as a gaussian state with a complex phase. The interested reader can find details concerning the case of a gaussian state with a real phase in appendix \ref{real} and we now tackle the issue of defining a physical state coming from a decoupled gaussian state with a complex phase.
Therefore, we consider a boundary state of the form,
\be \label{Factorizedgaussian}
\Psi_{\q}[j_{ab}]= \f{1}{\mathcal{N}}\prod_{a<b}e^{-\alpha (d_{j_{ab}}-d_{j_0})^2}e^{-i\Theta d_{j_{ab}}}
\ee
with $\mathcal{N}$ the normalization constant and $\alpha \in \C$. 
We now want to test this assumption using conditions (\ref{norm}) and (\ref{dyna}). These conditions lead to the two following equations on $\mathcal{N}$ and $\alpha$: 
\be \label{normapprox}
1=\sum_{j_{ab} }|\Psi_{\q}(j_{ab})|^2= \f{1}{(\mathcal{N}^2)^{10}}\sum_{\{j_{ab}\}} e^{-2\Re(\alpha)\sum_{a<b}(d_{j_{ab}}-d_{j_0})^2}
\ee
and,
\be \label{WWapprox}
1=\sum_{j_{ab} }K[d_{j_{ab}}] \Psi_{\q}(j_{ab})\simeq \frac{\mu}{\mathcal{N}^{10}}\sum_{\{j_{ab}\}}(\prod_{a<b}d_{j_{ab}})^\sigma P(d_{j_{ab}}) \sum_{\epsilon=\pm1} e^{-\Re(\alpha)\sum_{a<b}(d_{j_{ab}}-d_{j_0})^2+i\sum_{a<b}[d_{j_{ab}}(\epsilon \theta_{ab}-\Theta)-\Im(\alpha)(d_{j_{ab}}-d_{j_0})^2]}
\ee
The first equation corresponds to the normalization condition (\ref{norm}) for $\Psi_{\q}[j_{ab}]$. The second equation is the "Wheeler-deWitt" condition for $\Psi_{\q}[j_{ab}]$ in the simple case $K[j_{ab}]= \mu \left(\prod_{a<b}d_{j_{ab}}\right)^{\sigma}\tj$.  Solving them allow to determine uniquely $\mathcal{N}$ and $\alpha$ in term of the coefficients $\mu$ and $\rho$. 
The analysis is done in the large spin limit regime.
In equation (\ref{WWapprox}), we have already used the asymptotic formulae of the $\tj$-symbol. Moreover, we also replace $\alpha$ in (\ref{normapprox}) and (\ref{WWapprox}) by its asymptotic expression
\be
\alpha \sim \f{a+ib}{d_{j_0}^n}
\ee
with $a,\, b \in \R$ and $n$ restricted to belonging to $ ]0,2[$ in order that the asymptotic behavior of the  relative uncertainties of the area and angle on this state is correct.  Then solving the two obtained equations requires to distinguish three cases with respect to the value of the power $n \in \,]0,2[$. Keeping in mind that the standard choice of scaling for the Gaussian width $\alpha$ is $n=1$, the final result can be stated as:

\begin{proposition} The requirement on a factorized Gaussian state (\ref{Factorizedgaussian}) to be a physical state fixes the width of the Gaussian $\alpha$ for certain values of the power $n$:
\begin{enumerate}
\item $0<n<1$: the width of the Gaussian $\alpha$ is uniquely determined and the coefficient $\sigma$ is restricted. More precisely, \be \label{result1} \left\{ \tabl{l}{
\alpha \in \R_+, \\
a= \f{(\mu P)^{2/5} \pi}{2}, \; b=0\\
\sigma= \f14\left( \f95-n \right) \textrm{ and } \sigma>\f15,
} \right.
\ee
where $P$ is the numerical factor in front of the asymptotics of the $\{10j\}$-symbol.
\item $1<n<2$: there is no solution in this case.
\item the standard case $n=1$: A solution exists if and only if $$\sigma=\f15$$.
Then the value of $\alpha$ can be determined graphically or numerically in term of the value of $\mu P$. For example,
$$
a\simeq 1 \textrm{ and } b\simeq 1.99 \quad \textrm{ for } \mu P=1.01\,10^5.
$$
\end{enumerate}
\end{proposition}
Before proving this result, let us comment on the third case. This last case should be the most natural since the semi-classical state is then peaked on the same way on $\Theta$ encoding the extrinsic geometry of the boundary and on $d_{j_0}=A_0$ encoding the intrinsic geometry. However, the calculations do not look particularly more natural or transparent in that case. This leads to the conclusion that the choice of a decoupled gaussian state to define a physical state might be too simple and should be modified. A new proposition is given in the next subsection. Let us now give the proof of the results stated above.

\begin{proof}
In the large spin limit, the summation in equations (\ref{normapprox}) and (\ref{WWapprox}) can then be approximated with an integral  and we can write:
\be \label{intNorm}
1\simeq \f{1}{2\mathcal{N}^2} \int d(d_{j})e^{-2\Re(\alpha)(d_{j}-d_{j_0})^2}
\ee
and
\be \label{intDyna}
1\simeq \frac{\mu}{(2\mathcal{N})^{10}}\int \prod_{a<b}d[d_{j_{ab}}] (\prod_{a<b}d_{j_{ab}})^\sigma P(d_{j_{ab}}) \sum_{\epsilon=\pm1} e^{-\Re(\alpha)\sum_{a<b}(d_{j_{ab}}-d_{j_0})^2+i\sum_{a<b}[d_{j_{ab}}(\epsilon \theta_{ab}-\Theta)-\Im(\alpha)(d_{j_{ab}}-d_{j_0})^2]}
\ee
The first integral (\ref{intNorm}) is just a Gaussian integral, which can been integrated directly:
\be \label{equa1}
\mathcal{N}^2=\f{1}{2}\sqrt{\f{\pi}{2\Re(\alpha)}}
\ee
 given a first relation at the leading order between $\mathcal{N}$ and $\alpha$ .
 \\
 To evaluate the second integral (\ref{intDyna}) in the large spin limit, we first notice that the Gaussian implies
 $$\delta d_{j_{ab}}=d_{j_{ab}}-d_{j_0}\ll1\qquad \forall a<b,
 $$
  thus we can expand the Regge action around $d_{j_0}$:
 \be
 S_R[d_{j_{ab}}]=\sum_{a<b}d_{j_{ab}} \theta_{ab}(d_{j}) \simeq \sum_{a<b} d_{j_0} \Theta +\sum_{a<b} \f{\pp S_R}{\pp d_{j_{ab}}}|_{d_{j}=d_{j_0}} \delta d_{j_{ab}} +\f12\sum_{a<b, c<d} \f{\pp^2 S_R}{\pp d_{j_{ab}}d_{j_{cd}}}|_{d_{j}=d_{j_0}} \delta d_{j_{ab}} \delta d_{j_{cd}}
 \ee
The Schafli identity implies that
$$\f{\pp S_R}{\pp d_{j_{ab}}}|_{d_{j}=d_{j_0}} =\theta_{ab}(d_j)|_{d_{j}=d_{j_0}}= \Theta$$
since in the equilateral 4-simplex, all the dihedral angles are equal to $\Theta$. And we introduce the Hessian,
\be \label{hessian}
N^{ab}_{cd}= \f{\pp^2 S_R}{\pp d_{j_{ab}}d_{j_{cd}}}|_{d_{j}=d_{j_0}}=\f{\pp \theta_{ab}}{\pp d_{j_{cd}}}|_{d_{j}=d_{j_0}},
\ee
 then:
\be
 S_R[d_{j_{ab}}] \simeq 10d_{j_0} \Theta + \Theta \sum_{a<b} \delta d_{j_{ab}} +\f12 \sum_{a<b, c<d} N^{ab}_{cd}  \delta d_{j_{ab}} \delta d_{j_{cd}}=\Theta \sum_{a<b}  d_{j_{ab}} +\f12 \sum_{a<b, c<d} N^{ab}_{cd}  \delta d_{j_{ab}} \delta d_{j_{cd}}.
 \ee
We replace $ S_R[d_{j_{ab}}]$ by  this expansion in (\ref{intDyna}):
\be \tabl{ll}{
1 \simeq \frac{\mu}{(2\mathcal{N})^{10}}\int \prod_{a<b}d[d_{j_{ab}}] (\prod_{a<b}d_{j_{ab}})^\sigma P(d_{j_{ab}}) &[ e^{-2i\Theta\sum_{a<b}d_{j_{ab}}-\f{i}{2}\sum_{a<b,c<d}N^{ab}_{cd}  \delta d_{j_{ab}} \delta d_{j_{cd}}-\alpha \sum_{a<b}\delta d_{j_{ab}}^2}\\
&+ e^{\f{i}{2}\sum_{a<b,c<d}N^{ab}_{cd}  \delta d_{j_{ab}} \delta d_{j_{cd}}-\alpha \sum_{a<b}\delta d_{j_{ab}}^2}]
}\ee
The first exponential is a rapidly oscillating term in $d_{j_{ab}}$ which will vanish when we perform the integration over $d_{j_{ab}}$ so we only have to consider the second term. Moreover, at the leading order in $\delta d_{j_{ab}}$ we can replace $(\prod_{a<b}d_{j_{ab}})^\sigma P(d_{j_{ab}}) $  by $d_{j_{0}}^{10\sigma} P(d_{j_{0}})$ in the integral (\ref{intDyna}).

At this point, we recall that  $P(d_{j_{ab}})$ grows as $\zeta^{-9/2}$ when scaling all triangle areas $d_{j_{ab}}$ by $\zeta$. So we can write $P(d_{j_{ab}}=d_{j_0})\sim\f{P}{d_{j_0}^{9/2}}$ in the large $d_{j_0}$ limit, where $P>0$ is a constant. The precise value of $P$ can be computed analytically or determined numerically, but whose its explicit value does not affect the present discussion since it can be re-absorbed in the parameter $\mu$. Indeed, all the equations in the asymptotical regime will depend on $P$ only through the combination $\mu P$.

Then, once again we have to integrate a Gaussian integral:
\be \label{WWapprox2}
1\simeq \frac{\mu(d_{j_{0}})^{(10\sigma-9/2)} P}{(2\mathcal{N})^{10}}\int \prod_{a<b}d[\delta d_{j_{ab}}]  e^{-\delta d_{j_{ab}}M^{ab}_{cd}\delta d_{j_{cd}}}
\ee
where $M$ is a ten by ten matrix defined by
\be
M^{ab}_{cd}=\alpha \delta_{cd}^{ab} -iN^{ab}_{cd}\qquad a<b, \; c<d,
\ee
with  $\delta_{cd}^{ab}=1$ if the two couples of indices are the same and it vanishes otherwise;  $N^{ab}_{cd}=\f{\pp \theta_{ab}}{\pp d_{j_{cd}}}|_{d_{j}=d_{j_0}}$ was explicitly computed in \cite{graviton}\footnote{
$$N^{ab}_{cd}=\f{\sqrt{3}}{4\sqrt{5}d_{j_0}}\left(\tabl{cccccccccc}{
-4 & 7/2 & 7/2 & 7/2& 7/2 & 7/2 & 7/2 & -9 & -9 & -9\\
7/2 & -4 & 7/2 & 7/2 & 7/2 &-9 &-9 & 7/2 & 7/2& -9 \\
7/2 & 7/2 & -4 & 7/2 & -9 & 7/2 & -9 & 7/2 & -9 & 7/2\\
7/2 & 7/2 & 7/2 & -4 & -9 & -9 & 7/2 & -9 & 7/2 & 7/2\\
7/2 & 7/2 & -9 & -9 & -4 & 7/2 & 7/2 & 7/2 & 7/2 & -9 \\
7/2 & -9 & 7/2 & -9 & 7/2 & -4 & 7/2 & 7/2 & -9 & 7/2\\
7/2 & -9 & -9 & 7/2 & 7/2 & 7/2 & -4 & -9 & 7/2 & 7/2\\
-9 & 7/2 & 7/2 & -9 & 7/2 & 7/2 & -9 & -4 & 7/2 & 7/2 \\
-9 & 7/2 & -9 & 7/2 & 7/2 & -9 & 7/2 &7/2 & -4 & 7/2 \\
-9 & -9 & 7/2 & 7/2 & -9 & 7/2 & 7/2 & 7/2 & 7/2 & -4} \right)
$$}:
\be
N^{ab}_{cd}=\f{\tilde{N}^{ab}_{cd}}{d_{j_0}}
\ee
where $\tilde{N}$ is a ten by ten real symmetric constant matrix with all coefficients independent of $d_{j_0}$. In particular, $\tilde{N}^{ab}_{ab}=-\sqrt{\f35} \, \;\forall a<b$. Therefore, $M$ is a symmetric matrix with all its diagonal coefficients equal to $\alpha+i \sqrt{\f35}\f{1}{d_{j_0}}$.

At this stage we have to distinguish the three cases mentioned above. Assuming that $\alpha$ is of the form $\alpha=\f{a+ib}{d_{j_0}^n}$, with $a, b \in \R$, $n\in ]0,2[$. The three cases with respect to the power $n \in ]0,2[$ are: \begin{enumerate}
\item $0<n<1$ corresponds to $\alpha \gg \f{1}{d_{j_0}}$. In this case, we can neglect the terms of order $\f{1}{d_{j_0}}$ compared to $\alpha$.
\item $1<n<2$ corresponds to $ \alpha \ll \f{1}{d_{j_0}}$. In this case, $\alpha$ will be negligible with respect to the terms of the order $\f{1}{d_{j_0}}$.
\item $n=1$ corresponds to $\alpha \sim \f{1}{d_{j_0}}$. This case should be the most natural case since it peaks in the same way the triangle areas of the 4-simplex around the background value $A_0=d_{j_0}$ and the dihedral angles around the background value $\Theta$.
\end{enumerate}
Let us now separately study the three cases to solve equation (\ref{WWapprox2}).
\begin{enumerate}
\item {\bf The first case $0<n<1$} is the easiest one.
\\
Indeed in this case, the matrix $M\sim \f{a+ib}{d_{j_0}^n}+ \f{\tilde{N}}{d_{j_0}}$  can be approximated by a  ten by ten diagonal matrix: $M \simeq \alpha \id=(a+ib) \id$ since $\alpha \sim  \f{a+ib}{d_{j_0}^n} \gg \f{1}{d_{j_0}}$ and the ten integrals are then decoupled. Thus, we just have to compute a one-dimensional Gaussian integral:
\be \label{equa2-3}
\mathcal{N}= \f{(\mu P)^{1/10}}{2} (d_{j_{0}})^{(\sigma-9/20)} \int d[\delta d_j] e^{-\alpha (\delta d_{j})^2}= \f{(\mu P)^{1/10}}{2} d_{j_{0}}^{(\sigma-9/20)} \sqrt{\f{\pi}{\alpha}}
\ee
which gives a second equation for $\mathcal{N}$ and $\alpha$. Therefore using equation (\ref{equa1}), we obtain the following equation on $\alpha$:
\be
\f{1}{2}\sqrt{\f{\pi}{2\Re(\alpha)}}= \f{(\mu P)^{1/5}}{4} d_{j_{0}}^{(2\sigma-9/10)}\f{\pi}{\alpha}
\ee
Finally, expressing $\alpha$ under the form $\f{a+ib}{d_{j_0}^n }$, $0<n<1$), we get that:
\be
\left\{ \tabl{l}{
\alpha \in \R_+, \\
\sigma= \f14\left( \f95-n \right) \textrm{ and } \sigma>\f15, \\
a= \f{(\mu P)^{2/5} \pi}{2}, \quad b=0.
} \right.
\ee
In this case, $\alpha$ is real and positive. Furthermore, we get a condition on the normalization factor of the spinfoam vertex $\sigma > 1/5$.

\item {\bf The second case $1<n<2$} implies that $ \alpha \sim  \f{a+ib}{d_{j_0}^n} \ll \f{1}{d_{j_0}}$.
\\
In this case, the matrix $M$ reduces to $M=-iN$. We have to integrate: $\int d[\delta d_{j_{ab}}] \exp( i \sum \delta d_{j_{ab}} N^{ab}_{cd} \delta d_{j_{cd}})$. Recall that for a $m \times m $ symmetric invertible matrix $A$ with signature $\sigma(A)$ we have:
\be
\int_{\R^m}[dX_i] \exp \left[ i\left( \sum_{i,j} X_i A_{ij}X_j\right)\right]= e^{i\sigma(A) \f{\pi}{4}}\sqrt{\f{\pi^m}{|\det(A)|}}
\ee
In our case, $N$ is real, symmetric and its signature, which is the difference between the number of positive eigenvalues and the number of negative eigenvalues, is equal to $-2$. Therefore, $e^{i\sigma(N)\f{\pi}{4}}=-i$ and our Gaussian integral is an imaginary number which is not compatible with the first equation on $\mathcal{N}$ (\ref{equa1}). So, we cannot have $ \alpha \sim \f{a+ib}{d_{j_0}^n }$ with $1<n<2$.

\item {\bf The third case $n=1$} corresponds to $\alpha=\f{a+ib}{d_{j_0}}$.%
\\
 Then the determinant of $M$ is a complex number: its real part  and its imaginary part are polynomials of degree 10 and of arguments $a$ and $b$. We thus obtain a complex equation for $\mathcal{N}$ and $\alpha$:
\be
1\simeq \frac{\mu (d_{j_{0}})^{(10\sigma-9/2)} P}{(2\mathcal{N})^{10}}\int \prod_{a<b}d[\delta d_{j_{ab}}]  e^{-\delta d_{j_{ab}}M^{ab}_{cd}\delta d_{j_{cd}}}=\frac{\mu (d_{j_{0}})^{(10\sigma-9/2)} P}{(2\mathcal{N})^{10}} \sqrt{\frac{\pi^{10}}{ \det(M)}}
\ee
Combining this equation with the first equation (\ref{equa1}) that we already have on $\mathcal{N}$ and $\Re(\alpha)=a$ we get a complex equation on $a$ and $b$:
\be \label{equa2-1}
2^5\left(\f{d_{j_0}\pi}{2a}\right)^{5/2}-\mu (d_{j_{0}})^{(10\sigma-9/2+5)} P\sqrt{\frac{\pi^{10}}{ \det(\tilde{M}(a,b))}}=0
\ee
with $\tilde{M}=\f{M}{d_{j_0}}$. This equation implies a condition on the normalization factor of the spinfoam vertex: $$
\sigma=\f15.
$$
 We can then solve numerically the previous equation by plotting a 3d graph representing the square of the norm of the complex number given by the left-hand side of the previous equation (\ref{equa2-1}) in terms of $a$ and $b$ using Maple (see graph fig. \ref{graphsurface}). For example, the value of this norm vanishes for $$
 a\simeq 1 \textrm{ and } b\simeq 1.99 \quad \textrm{ for } \mu P=1.01\,10^5.
 $$
  Therefore, there exists a specific value $\alpha$ for which $\Psi_{\q}$ is a physical state. However, whereas this case for which $\Psi_{\q}$ is peaked in the same way around the intrinsic and extrinsic geometry $\q$ should be the most natural, it is quite complicated.
\end{enumerate} \qed
\end{proof}

\begin{figure}[h]
\begin{center}
\includegraphics[height=50mm]{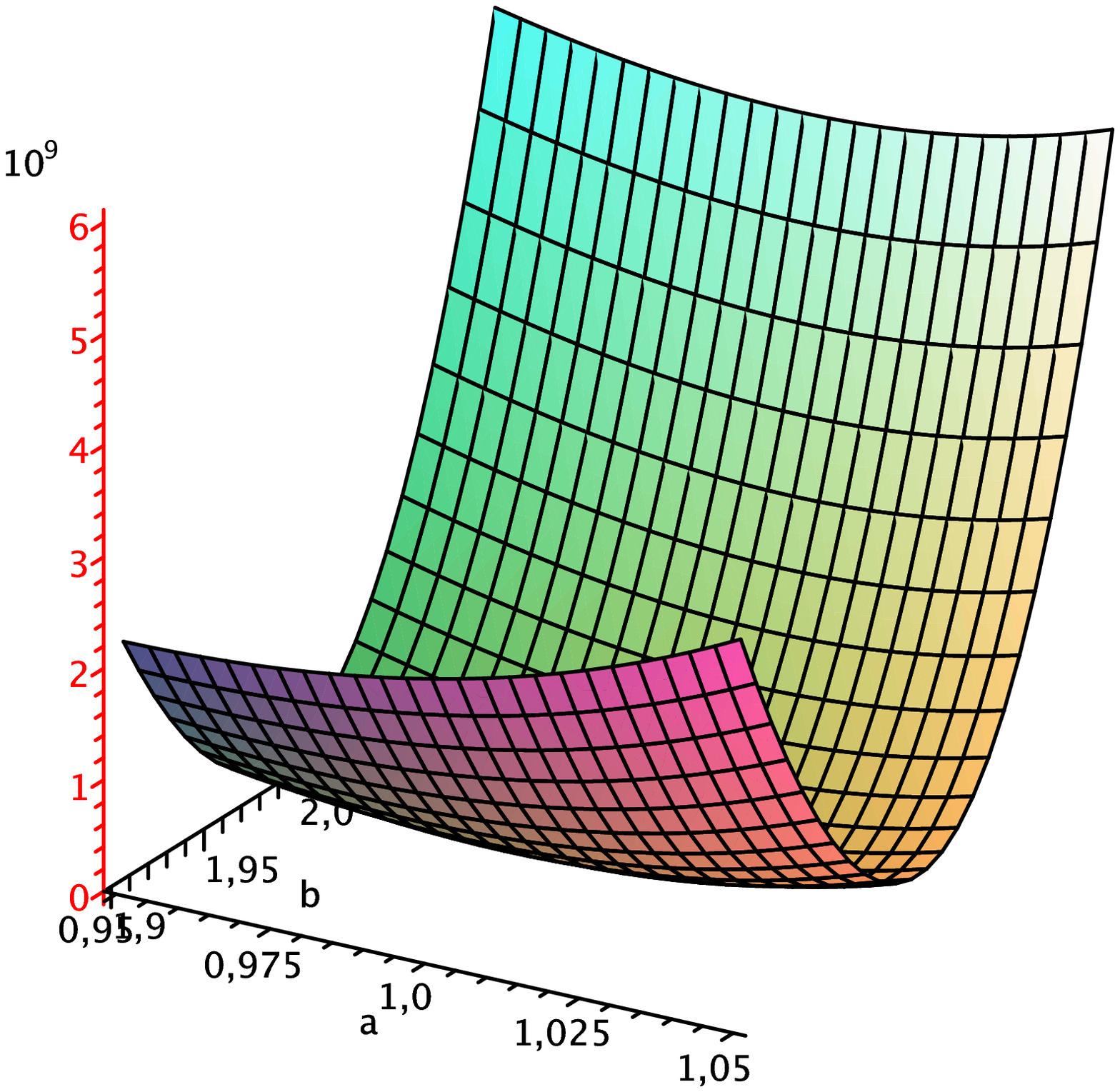}
\hspace{20mm}
\includegraphics[height=40mm]{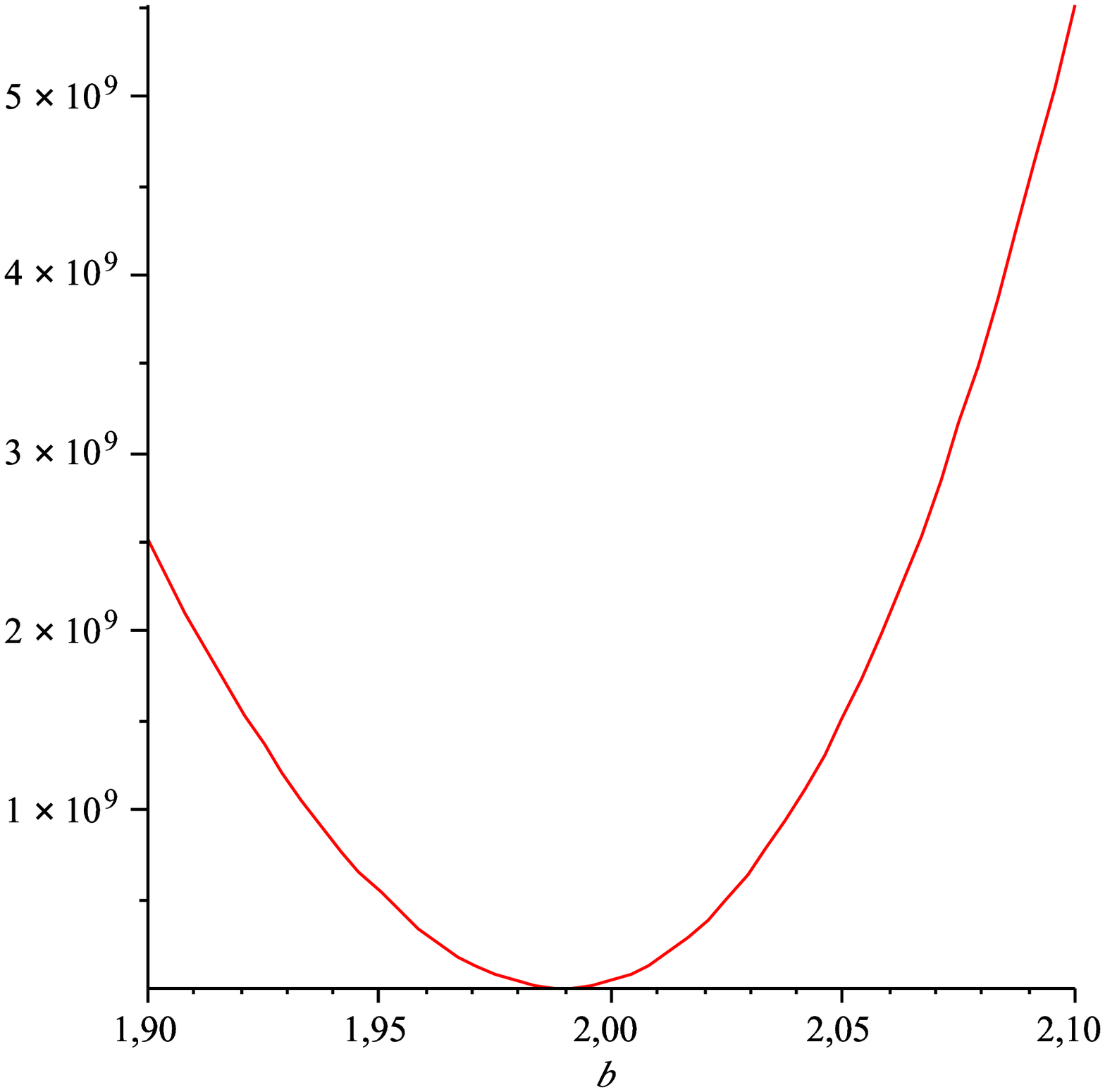}
\caption{We plot the norm squared of the left hand side of eqn \Ref{equa2-1} for $\mu P=1.01\,10^5$ in term of $a$ and $b$. The second graph is a slice of the 3d graph for fixed $a=1$ and variable $b$. \label{graphsurface}}
\end{center}
\end{figure}

The alternative is then to work with a more complication semi-classical state. Indeed although a factorized gaussian wave-packet  is up to now the only one which has allowed to perform numerical simulations, the previous analysis seems to show that it is too simple to catch all the features of a physical state. We show in the next section that  a tensorial Gaussian state is more adapted to describe a physical state. Indeed we will see that such a state allows to compensate the imaginary part which comes from the second derivative of the Regge action and given by the matrix $iN$ and consequently to simplify the resolution of this third case studied above. This is what's commonly expected from the more general perspective of Regge calculus \cite{bianca}.

\subsection{The Coupled Gaussian Ansatz}

To go beyond the decoupled Gaussian ansatz, our new assumption is to consider a boundary state of the form:
\be \label{coupledGaussian}
\psi_q[d_{j_{ab}}]= \f{1}{\mathcal{N}}e^{-\sum_{a<b,c<d}\alpha_{cd}^{ab} (d_{j_{ab}}-d_{j_0})(d_{j_{cd}}-d_{j_0})}e^{i\Theta \sum_{a<b}d_{j_{ab}}}
\ee
where $\alpha$ is now a ten by ten complex matrix. Furthermore we choose
\be \label{coupledWidth}
\alpha_{cd}^{ab}= \beta(d_{j_0}) \delta^{ab}_{cd} + iN_{cd}^{ab}
\ee
 where $\beta \in \R$ and the imaginary part of $\alpha$ is  now  the conjugate variables of the dihedral angles of the tetrahedron in the semi-classical regime introduced in (\ref{hessian}):$$
 N^{ab}_{cd}=\f{\pp \theta_{ab}}{\pp d_{j_{cd}}}|_{d_{j}=d_{j_0}}.
 $$
$N^{ab}_{cd}$ is the Hessian of $S_R$ and we will see that it is this choice which allows to simplify the construction of a physical semi-classical state peaked in the same way on the extrinsic and extrinsic geometry of the 3d boundary. We recall that $N$ depends on $d_{j_0}$ such that $N=\f{\tilde{N}}{d_{j_0}}$ with $\tilde{N}$ a matrix with constant coefficients.
\begin{proposition}
For $0<n\leq 1,$ we take as ansatz for the width $\beta =\f{a}{d_{j_0}^n}$ (at least at leading order for large $j_0$) of the Gaussian state (\ref{coupledGaussian}). This defines the real part of the matrix $\alpha$ through \eqref{coupledWidth}. Then the width is uniquely defined and the coefficient $\sigma$ is restricted. More precisely,
\be \label{resultcoupled1} \left\{ \tabl{l}{
\beta \in \R_+, \\
a= \f{(\mu P)^{2/5} \pi}{2}\\
\sigma= \f14\left( \f95-n \right) \textrm{ and } \sigma\geq\f15,
} \right.
\ee
\end{proposition}

Therefore, the case $n=1$ appears now in the continuity of the case $0<n<1$. This can be considered as an improvement compared to the previous case of a factorized Gaussian semi-classical state. However, we have not solved the case $1<n<2$, which admitted no solution in the factorized Gaussian state. The resolution of this case needs the knowledge of the next-to-leading order of the asymptotic expansion of the $\{10j\}$-symbol, which hasn't yet been computed.

\begin{proof}
Regarding the asymptotic behavior of $\beta$, the real part of the matrix $\alpha$, we distinguish the three same cases as in the decoupled gaussian state case.

\begin{enumerate}
\item  The result of the {\bf first case}, $\beta \sim \f{a}{d_{j_0}^n}$  with $0<n<1$, is not modified. therefore, {\bf $\beta \gg \f{1}{d_{j_0}}$} and the matrix $M$ can be approximated by a ten by ten diagonal matrix $M\simeq \beta \, \id$ and the state is again factorized and we obtain the same result as in (\ref{result1}):
\be \label{result2} \left\{ \tabl{l}{
\beta \in \R_+ \\
a= \f{(\mu P)^{2/5} \pi}{2}\\
\sigma= \f14\left( \f95-n \right) \textrm{ and } \sigma> \f15
} \right.
\ee
\item The {\bf second case} is when $1<n<2$ in $\beta \sim \f{a}{d_{j_0}^n}$ ($a\in \R$). Since {\bf $\beta \ll \f{1}{d_{j_0}}$} in this case, the real part of $\alpha$ is negligible compare to its imaginary part.  Consequently since this imaginary part compensates the second derivative of the Regge action in the exponential of the second condition (\ref{dyna}), we should go to the next order in $\delta d_j$. To compute this expansion in $\delta d_j$ we need to know the next-to-leading order of the asymptotic expansion of the $\tj$-symbol. We can thus say nothing now in this case for the moment.

\item The {\bf third case}, {\bf $\beta \sim \f{a}{d_{j_0}}$} ($a \in \R$), is greatly simplified compared to the equivalent case for the decoupled gaussian wave-packet. Indeed, applying condition (\ref{norm}) and condition (\ref{dyna}) to this state, we obtain the two following equations for $\mathcal{N}$ and $a$:
\be \left\{ \tabl{l}{
1\simeq \f{1}{2^{10}\mathcal{N}^2}\left(\int d[\delta d_{j}] e^{-2\f{a}{d_{j_0}}(\delta d_{j})^2} \right)^{10}= \f{1}{2^{10}\mathcal{N}^2} \left(\frac{\pi d_{j_0}}{2a} \right)^5\\
1 \simeq \f{d_{j_0}^{10\sigma-9/2}\mu P}{2^{10}\mathcal{N}}\left(\int d[\delta d_j] e^{-\f{a}{d_{j_0}}(\delta d_j)^2}\right)^{10}=  \f{d_{j_0}^{10\sigma-9/2}\mu P}{2^{10}\mathcal{N}}\left(\frac{\pi d_{j_0}}{a} \right)^5
} \right.
\ee
where we use to compute the second equation the same analysis as in the factorized gaussian state case.  The key element which simplifies everything is that the second derivative of the Regge action added to the imaginary part of $\alpha$ is null since $\alpha$ imaginary part is the Hessian. This system of equations can then be simplified in a single equation for $a$:
\be
\f{1}{2^{10}}\left( \f{\pi d_{j_0}}{2a} \right)^5 =\f{d_{j_0}^{(20\sigma-9)}(\mu P)^2}{2^{20}} \left( \f{\pi d_{j_0}}{a}\right)^{10}
\ee
which implies a condition on the normalization factor of the $\tj$-symbol in the spinfoam vertex:
\be
\sigma= \f{1}{5}
\ee
 and the unique solution for $a$:
 \be
 a= \f{\pi (\mu P)^{2/5}}{2}.
 \ee
 Therefore, we now get the same value of $a$ in this case as in the first case. Moreover, $\sigma$ is now uniquely determined since  $n=1$.
\end{enumerate}
\end{proof}
This analysis on the consequences of conditions (\ref{norm}) and (\ref{dyna}) on a coupled gaussian wave-packet (\ref{coupledGaussian}) 
is therefore more convicting on the ability of such a state to capture a true physical rather than a too simple state defined  as a factorized gaussian state. Indeed, the width $\alpha$ of the coupled gaussian state is now uniquely determined by the requirement of a normalized physical state, as expected from the 3d case \cite{physical3d}.

\section*{Conclusion and Outlook}
The analysis done shows that it is possible to determine a physical semi-classical state for the Barrett-Crane model in the case of the simplest triangulation given by a 4-simplex. The requirement of a normalized physical state determines uniquely the Gaussian width (in term of the precise normalization of the $\{10j\}$ vertex amplitude). This analysis shows also that we cannot take a too simple semi-classical state. For e.g. we have seen that a decoupled gaussian state which is peaked on the geometry of an equilateral 4-simplex does not seem to be able to capture a true physical state. A coupled gaussian state appears as defining more naturally a physical semi-classical state. 
\smallskip
We recall that in all our calculations, we have used $K_1[s]$ the lowest order term in the expansion of $K[s]$ in the group field theory coupling constant $\lambda$  and that  $s$ symbolizes the 4-simplex boundary graph. The total propagator kernel $K[s]$ is given by  $K[s]=\sum_{V=1}^{\infty} \lambda^{V}K_V[s]$. This operator should describe a unitary "evolution". In the spin foam framework the notion of evolution is not well-defined, 
however we can require the following normalization condition on $K[s]$
\be
K[s]\bar{K}[s]=(K[s])^2=1
\ee
that describes the process of creation of a  4-simplex starting from a null 4-volume following by the  annihilation of this 4-simplex into a null 4-volume again. This condition also says that $K$ is by definition a physical state. Using the $\lambda$ expansion of  $K[s]=\sum_{V=1}^{\infty} \lambda^{V}K_V[s]$, we can expand this normalization condition in power of $\lambda$
\be
\sum_{V=1}^\infty \lambda^{V}\sum_{V_1, V_2/V_1+V_2=V+1}K_{V_1}K_{V_2}=1
\ee
which simplifies  at the leading order in $\lambda$ in
\be
\lambda^2K_1[s]^2=1
\ee
meaning that approximatively $\lambda K_1[s]$ is a physical state and  introducing a new constraint on $\lambda K_1[s]=  \lambda\mu \prod_{a<b}d_{j_{ab}}^{\sigma}\tj$. It is this new constraint which allows to determine $\mu$:
\be
(\lambda\mu)^2=\f{1}{\left(\prod_{a<b}d_{j_{ab}}\right)^{2\sigma}\tj^2}
\ee
where $\sigma \geq 1/5$ in order that $\mu$ is finite. This restriction on the domain of validity of $\sigma$ is consistent with the results obtained previously.
\smallskip
Since up to now we have only considered the lowest order term in the expansion of $K[s]$ in  $\lambda$, we have in fact fixed  the simplest bulk triangulation and taken into account only a finite number of degrees of freedom.
The next step would be, keeping the same boundary spin network $s$, to explicitly write the next terms: $K_2$, $K_3$,... of the $\lambda$-expansion of $K[s]=\sum_{V=1}^{\infty} \lambda^{V}K_V[s]$. This is not obvious.  A priori,  $K_4$ could be determined from $K_1$ by doing a 5-1 move. Otherwise, from an effective field theory point of view we can  write $K[s]=K_1[s]+ k[s]$ where $k$ takes into account all the possible counter-terms and therefore all the possible bulk geometries for a given boundary spin network. This is equivalent to look at a coarse-grained lattice in which all vertices can be considered as shrunk to a single effective vertex. $k[s]$ is then the weight associated to this effective vertex. In this weight $k[s]$, we expect to get a term proportional to the original Barrett-Crane spinfoam vertex amplitude, $\epsilon \tj$, which would come from the contribution of the flat 4-simplex. Moreover, we expect additional contributions representing curved 4-simplex. These extra terms can be written in a condensed way as $\rho  \int dG\, P(G) \,\Gamma_{10j}(G)$, where $\Gamma_{10j}(G)$ is the  spin network functional, on the 4-simplex boundary, evaluated on the non-trivial holonomies $G$. The flat 4-simplex is then simply the term
$\Gamma_{10j}(\id)=\tj$, and $P(G)$ is the measure factor giving the relative weight of every curved configuration.
Other terms will certainly appear in a more precise analysis.
The issue is then to determine which terms of $k[s]$ contribute in the requirement of a physical state for $\psi_q$; that is, we should understand how the condition $\sum_{j_{ab}} K[d_{j_{ab}}] \psi(d_{j_{ab}})=1$ is explicitly modified by the term $k[d_{j_{ab}}]$ of $K[d_{j_{ab}}] $. We would expect that the requirement for the boundary state to be a physical state selects the bulk triangulation. From this perspective, the choice of the phase seems crucial: indeed it is the choice of the flat dihedral angles in the phase factor of the boundary state that is used to select the flat 4-simplex at leading order. The choice of  a different phase for $\Psi_q$ should select another bulk triangulation $\Gamma$, which might correspond a non-trivial space-time curvature. A detailed investigation is required to confirm this rough analysis and to really understand the dominant terms in the computation of the two-point
 function of the gravitational field.
\smallskip
Up to now, we have discussed the case of the simplest 4d triangulation defined by a single 4-simplex. We should go beyond and study more general backgrounds defined on more complex triangulations. We would then expect a more complicated dependence on the boundary spins of both the Gaussian matrix and the phase factor, as happens in Regge calculus (see e.g. \cite{bianca}). To deal with such more complex backgrounds, we would however need to develop more powerful analytical tools to probe the spinfoam amplitudes and their asymptotics. We nevertheless hope to tackle this issue soon.
\smallskip
Finally, another step would be to deal with the EPRL-FK models instead of the Barrett-Crane model to define the propagator kernel $K[s]$. Since the EPRL-FK vertex has also for asymptotic the exponential of the Regge action, we are expected to get similar results in the restricted case of a single 4-simplex. However, the behavior under renormalization should be quite different since the space of intertwiners in the EPRL-FK model introduces new degrees of freedom in the 3d space geometry compared to the Barrett-Crane geometry which is describe by a unique intertwiner and  seems to have too many frozen degrees of freedom. These degrees of freedom will have to be taken into account when gluing 4-simplices together.



\appendix

\section{Physical States with a Real Phase} \label{real}
We will discuss here the case already mentioned of the Bessel-based factorized boundary state:
\be
\psi_q[j_{ab}]= \prod_{a<b}\phi_B(j_{ab})
\ee
where
\be
\phi_B(j)= \f{I_{|d_j-d_{j_0}|}(\f{d_{j_0}}{\alpha})-I_{d_j+d_{j_0}}(\f{d_{j_0}}{\alpha})}{\sqrt{I_0(\f{2d_{j_0}}{\alpha})-I_{2d_{j_0}+1}(\f{2d_{j_0}}{\alpha})}} \cos\left(d_j\Theta\right)
\ee
We are interested in the large spin limit regime and its $d_{j_0} \rightarrow \infty$ limit behaves as a Gaussian peaked around $d_{j_0}$:
\be
\phi_B(j)\simeq \left( \f{\alpha}{ d_{j_0}\pi }\right)^{1/4} \exp\left[ -\f{\alpha}{2 d_{j_0}}(d_j-d_{j_0})^2\right] \cos (d_j \Theta)
\ee
The difference with the case studied previously of a Gaussian peaked around $d_{j_0}$ is the phase which is real here.
We recall that the factorized boundary state assumption has been made in order to perform numerical simulations and the choice of a real phase has been done in the work concerning the area correlator to turn it into an exact group integral \cite{simone1} and perform exact analytical computations.

We therefore consider now a factorized Gaussian state with a real phase:
\be
\psi_q[d_{j_{ab}}] \f{1}{\mathcal{N}}\prod_{a<b} \exp(-\alpha (d_{j_{ab}}-d_{j_0})^2) \cos(d_{j_0}\Theta)
\ee
 to see if the conditions necessary to have a physical state are modified.  Conditions (\ref{norm}) and (\ref{dyna}) give the two following equations for $\mathcal{N}$ and $\alpha$.
\be
1=\sum_{j_{ab} }|\psi(j_{ab})|^2= \f{1}{(\mathcal{N}^2)^{10}}\sum_{\{j_{ab}\}} e^{-2\Re(\alpha)\sum_{a<b}(d_{j_{ab}}-d_{j_0})^2} \prod_{a<b} \left(\f{1+ \cos(2d_{j_{ab}}\Theta)}{2} \right)
\ee
and,
\be
1=\sum_{j_{ab} }\tj \psi(j_{ab})\simeq \frac{1}{\mathcal{N}^{10}}\sum_{\{j_{ab}\}}(\prod_{a<b}d_{j_{ab}})^\sigma P(d_{j_{ab}}) \sum_{\epsilon=\pm1;\, \eta=\pm1} e^{-\Re(\alpha)\sum_{a<b}(d_{j_{ab}}-d_{j_0})^2+i\sum_{a<b}[d_{j_{ab}}(\epsilon \theta_{ab}+\eta \Theta)-\Im(\alpha)(d_{j_{ab}}-d_{j_0})^2]}
\ee
where we have already used the asymptotic formulae of the $\tj$-symbol. In the large spin limit, the summation in the two previous equations can then be approximated with an integral  and we can write:
\be \label{intNorm2}
1\simeq \f{1}{2\mathcal{N}^2} \int d(d_{j})e^{-2\Re(\alpha)(d_{j}-d_{j_0})^2} \left[\f12 +\f14 e^{i2d_j \Theta}+ \f14 e^{-i2d_j\Theta} \right]
\ee
and
\be \label{intDyna2}
1\simeq \frac{1}{(2\mathcal{N})^{10}}\int \prod_{a<b}d[d_{j_{ab}}] (\prod_{a<b}d_{j_{ab}})^\sigma P(d_{j_{ab}}) \sum_{\epsilon=\pm1;\, \eta=\pm} e^{-\Re(\alpha)\sum_{a<b}(d_{j_{ab}}-d_{j_0})^2+i\sum_{a<b}[d_{j_{ab}}(\epsilon \theta_{ab}+\eta\Theta)-\Im(\alpha)(d_{j_{ab}}-d_{j_0})^2]}
\ee
The first integral (\ref{intNorm2}) after some changes of variables is just three Gaussian integrals, which can been integrated directly:
\be
\mathcal{N}^2=\f{1}{4}\sqrt{\f{\pi}{2\Re(\alpha)}}\left[1+e^{-\f{\Theta^2}{2\Re(\alpha)}} \cos(d_{j_0} \Theta) \right]
\ee
but we recall that $\alpha \propto \f{1}{d_{j_0}^n}$ with $n \in ] 0,2 [$ from equations (\ref{uncertainties}) and therefore in the large spin limit the second term of the right-hand side of this relation vanishes and we obtain the same relation at the leading order between $\mathcal{N}$ and $\alpha$ as in the case of a Gaussian state with an imaginary phase (up to a factor $1/2$). That is:
\be
\mathcal{N}^2=\f{1}{4}\sqrt{\f{\pi}{2\Re(\alpha)}}
\ee
 To evaluate the second integral (\ref{intDyna2}) in the large spin limit we expand as previously the Regge action around $d_{j_0}$:
\be
 S_R[d_{j_{ab}}] \simeq \Theta \sum_{a<b}  d_{j_{ab}} +\f12 \sum_{a<b, c<d} N^{ab}_{cd}  \delta d_{j_{ab}} \delta d_{j_{cd}}.
 \ee
 where  $$N^{ab}_{cd}= \f{\pp^2 S_R}{\pp d_{j_{ab}}d_{j_{cd}}}|_{d_{j}=d_{j_0}}=\f{\pp \theta_{ab}}{\pp d_{j_{cd}}}|_{d_{j}=d_{j_0}}=\f{\sqrt{3}}{4\sqrt{5}d_{j_0}}\left(\tabl{cccccccccc}{
-4 & 7/2 & 7/2 & 7/2& 7/2 & 7/2 & 7/2 & -9 & -9 & -9\\
7/2 & -4 & 7/2 & 7/2 & 7/2 &-9 &-9 & 7/2 & 7/2& -9 \\
7/2 & 7/2 & -4 & 7/2 & -9 & 7/2 & -9 & 7/2 & -9 & 7/2\\
7/2 & 7/2 & 7/2 & -4 & -9 & -9 & 7/2 & -9 & 7/2 & 7/2\\
7/2 & 7/2 & -9 & -9 & -4 & 7/2 & 7/2 & 7/2 & 7/2 & -9 \\
7/2 & -9 & 7/2 & -9 & 7/2 & -4 & 7/2 & 7/2 & -9 & 7/2\\
7/2 & -9 & -9 & 7/2 & 7/2 & 7/2 & -4 & -9 & 7/2 & 7/2\\
-9 & 7/2 & 7/2 & -9 & 7/2 & 7/2 & -9 & -4 & 7/2 & 7/2 \\
-9 & 7/2 & -9 & 7/2 & 7/2 & -9 & 7/2 &7/2 & -4 & 7/2 \\
-9 & -9 & 7/2 & 7/2 & -9 & 7/2 & 7/2 & 7/2 & 7/2 & -4} \right).$$
 And using once again the argument that rapidly oscillating term in $d_{j_{ab}}$ will vanish when preforming the integration over $d_{j_{ab}}$, we have to consider only two terms in the sum (\ref{intDyna2}) which are two Gaussian integral:
\be
1\simeq \frac{(d_{j_{0}})^{(10\sigma-9/2)} P}{(2\mathcal{N})^{10}}\int \prod_{a<b}d[\delta d_{j_{ab}}] \left[ e^{-\delta d_{j_{ab}}M^{ab}_{cd}\delta d_{j_{cd}}} + e^{-\delta d_{j_{ab}}Q^{ab}_{cd}\delta d_{j_{cd}}}  \right]
\ee
where $M$  and $Q$ are both  ten by ten matrices defined by: $M^{ab}_{cd}=\alpha \delta_{cd}^{ab} -iN^{ab}_{cd}$ and $Q^{ab}_{cd}=\alpha \delta_{cd}^{ab} +iN^{ab}_{cd}$ with $a<b$, $c<d$ and $\delta_{cd}^{ab}=1$ if the two couples of indices are the same and it vanishes otherwise. Therefore, $M$ is a symmetric matrix with all its diagonal coefficients equal to $\alpha+i\sqrt{\f35}\f{1}{d_{j_0}}$ and $Q$ is a symmetric matrix with all its diagonal coefficients equal to $\alpha-i \sqrt{\f35}\f{1}{d_{j_0}}$. Then the analysis is the same as the one done in the case of a Gaussian state with an imaginary phase;  we have three case to consider:\begin{enumerate}
\item $\alpha \gg \f{1}{d_{j_0}}$, that is we consider here that $\alpha$ is proportional to $d_{j_0}^{-n}$ with $0<n<1$. We have then that $M \simeq \alpha Id$ and $Q\simeq \alpha Id$ and once again we write $\alpha=\f{a}{d_{j_0}^n}$, then we get the same result as in the case of the imaginary phase (up to some factors $2$ in $a$) (see equations \ref{result1}), that is: $\left\{ \tabl{l}{
\alpha \in \R \\
\sigma= \f14\left( \f95-n \right) \textrm{ and } \sigma>\f15 \\
a= 32P^4 \pi
} \right.$
\item $ \alpha \ll \f{1}{d_{j_0}}$, that is we consider that $\alpha$ is proportional to $d_{j_0}^{-n}$ with $1<n<2$. We have to integrate $\int d[\delta d_{j_{ab}}] \exp( i \sum \delta d_{j_{ab}} N^{ab}_{cd} \delta d_{j_{cd}})+ \exp(- i \sum \delta d_{j_{ab}} N^{ab}_{cd} \delta d_{j_{cd}})$. The signature of the matrix $N$ is equal to $-2$; therefore, $e^{i \sigma(N)\f{\pi}{4}}=-i=-e^{i \sigma(-N)\f{\pi}{4}}$ and then the previous integral is null so in this case we also cannot have $\alpha=\f{a}{d^n_{j_0}}$ with $1<n<2$.
\item $\alpha \sim \f{1}{d_{j_0}}$: it is the most natural case which peaks in the same way the triangle areas of the 4-simplex around the background value $A_0=d_{j_0}$ and the dihedral angles around the background value $\Theta$. But once again due to the fact we have chosen a factorized boundary state, this case is very complicated although the imaginary parts of the determinant of $M$ and of the determinant of $N$ will compensate.  This confirms that a factorized boundary state is not the most natural to capture a physical state. And with a real phase it is not even possible to consider a tensorial state which could compensate $iN$ and $-iN$ like we did in the case of the imaginary phase.

\end{enumerate}


\end{document}